\documentclass[review]{elsarticle}

\usepackage{lineno,hyperref}
\modulolinenumbers[5]

\journal{Materials Research Bulletin}

\usepackage{url}  
\usepackage{graphicx} 
\usepackage{revsymb}
\usepackage{bm}
\usepackage{textcomp}
\usepackage{revsymb}
\usepackage{amssymb}
\usepackage{ifsym}
\usepackage{wasysym}
\usepackage{txfonts}
\usepackage{hyperref}
\usepackage{tipa}

\begin{document}
	
	\begin{frontmatter}

		
	\title{CdZnSSe crystals synthesized in silicate glass: structure, cathodoluminescence, band gap, discovery in historical glass, and possible applications in contemporary technology\tnoteref{funding,license}}
		

		\author[GOSNIIR]{Tatyana~V.~Yuryeva\corref{correspondingauthor}} 
		\ead{yuryevatv@gosniir.ru}

		\author[MSU,PI,UEF]{Sergey~A.~Malykhin\corref{malykhin_grants}}
		
		\author[TESCAN]{Andrey~A.~Kudryavtsev}
		
		\author[RF_CFS]{Ilya~B.~Afanasyev}
		
		\author[GOSNIIR]{Irina~F.~Kadikova} 
		
		
		\author[GPI]{Vladimir~A.~Yuryev\corref{correspondingauthor}}
		\ead{vyuryev@kapella.gpi.ru}
		
		\address[GOSNIIR]{The State Research Institute for Restoration, Building 1, 44 Gastello Street, Moscow 107114, Russia}
		
		\address[MSU]{Faculty of Physics, M.\,V.\,Lomonosov Moscow State University, Leninskie Gory, Moscow 119991, Russia} 
		
		\address[PI]{P.\,N.\,Lebedev Physical Institute of the Russian Academy of Sciences, 53 Leninsky Avenue, Moscow 119991, Russia} 
		
		\address[UEF]{Department of Physics and Mathematics, University of Eastern Finland, Yliopistokatu 7, Joensuu 80100, Finland}
		
		\address[TESCAN]{TESCAN Ltd., PO Box 24, Office 212, 11A Grazhdansky Avenue, Saint-Petersburg 195220, Russia}
		
		\address[RF_CFS]{The Russian Federal Center of Forensic Science of the Ministry of Justice, Bldg.~2, 13 Khokhlovskiy Sidestreet, Moscow 109028, Russia}
		
		
		\address[GPI]{A.\,M.\,Prokhorov General Physics Institute of the Russian Academy of Sciences, 38 Vavilov Street, Moscow 119991, Russia}
		

		\date{September 27, 2019}

\cortext[correspondingauthor]{Corresponding author.}
\cortext[malykhin_grants]{Grantee of the ``BASIS'' Foundation and the Academy of Finland [grant number 298298].}
\tnotetext[funding]{This work was supported by the Russian Science Foundation [grant number 16-18-10366] and the Russian Foundation for Basic Research [grant number 18-312-00145].}
\tnotetext[license]{\textcopyright\,2019.~This manuscript version is made available under the CC~BY-NC-ND~\textbf{}4.0 license {http://creativecommons.org/licenses/by-nc-nd/4.0/}.}

\begin{abstract}

The article presents investigations of CdZnSSe crystals synthesized in silicate glass melt. 
Zn-rich glass manufactured in the 19th century has been found to contain CdZnSSe crystals of hexagonal crystal system exhibiting intense band-edge cathodoluminescence. 
The exciton luminescence peak of these crystals ranges 2.00 to 2.03~eV at 300~K and 2.02 to 2.06~eV at 80~K. 
Its shifts in individual crystals are attributed to minor variations of their composition. The value of the band gap derivative $dE_{\rm g}/dT$ lies in the range from $-2.5\times 10^{-4}$ to 0 eV/K in the interval from 80 to 300~K. 
CdZnSSe crystals in glass demonstrate high temperature and temporal stability. 
The glass-crystal composite on their basis is resistant to the electron-beam irradiation and long-term weathering. 
Possible applications of this composite in modern technologies, and processes and components that might be used for making glass stained with CdZnSSe in the past are also discussed.

\end{abstract}

\begin{keyword}
	\texttt{ A. glasses \sep A. semiconductors \sep B. luminescence \sep B. chemical synthesis \sep C. electron microscopy}
\end{keyword}

\end{frontmatter}


\begin{figure}[h!]
	\includegraphics[width=\textwidth]{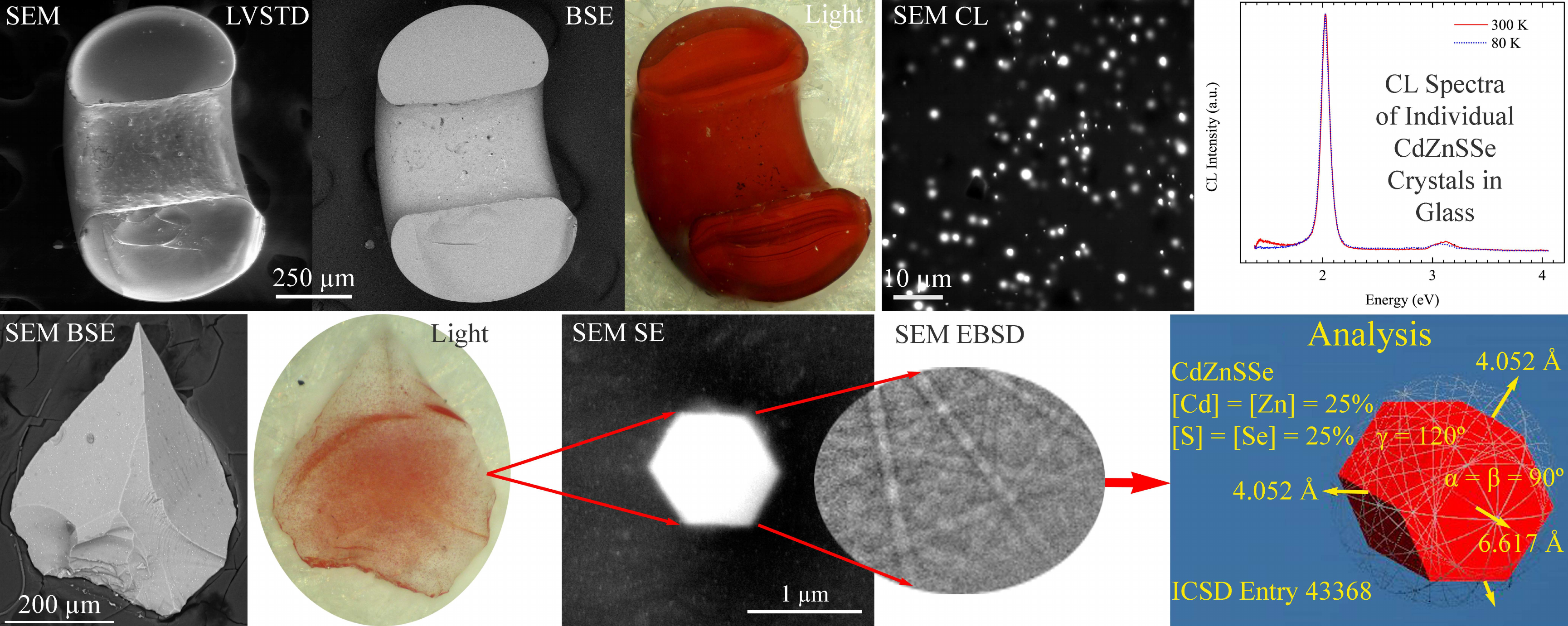}
\end{figure}

\section*{Highlights}
\begin{itemize}
	\small
\item
The article presents studies of CdZnSSe crystals synthesized in silicate glass melt
\item
Hexagonal CdZnSSe crystals were discovered in 19th century Zn-rich red bead glass
\item
CdZnSSe crystals in glass exhibit intense exciton luminescence peaked around 600 nm
\item
CdZnSSe crystals in glass demonstrate high temperature and temporal stability
\item
CdZnSSe-glass composite is resistant to electron beams and long-term weathering
\end{itemize}

\newpage

\section{Introduction}
\label{intr}

Investigation of physical, especially optical properties of crystallites in glass, in particular in silicate glass, as well as the properties of the glass-crystal composite as a whole, is a promising area of researches paving the way for numerous practical applications of such materials.
Highly luminescent glasses forming due to the synthesis of luminescent crystallites directly in the process of glass melting are of special interest due to simplicity of their shaping and processing either using standard technologies of glassware making or using the technology of 3D printing \cite{3D_printing_ceramics}.
For instance, glass films or fibers containing luminescent micro crystallites or nano-crystalline colloidal quantum dots may be promising materials for light-emitting devices.
They may be useful in the lighting industry, say, for the production of light emitting diodes by rare-earth-free phosphor-in-glass technology \cite{PL_Glasses}.
Colloidal staining of glass using luminescent crystallites emitting visible light is also attractive for making artworks of glass \cite{PL_Glasses,GLASS_AN_USSR,Art_items_glass,Luminescent_glasses_art} or in the architecture \cite{PL_Glasses}.

Among numerous crystalline compounds appropriate to formation of light emitting crystallites in glass, as well as other light emitting structures, wide-bandgap II--VI semiconductors based on Zn, Cd, Se and S attract maybe the highest attention due to the possibility to vary the band gap in a wide range from 3.8 to 1.74 eV by varying their composition and hence to move the peak of the exciton-related emission over the spectral range from ultraviolet (ZnS) to deep red (CdSe) \cite{Pankove,II-VI_compounds}. 
The advantage of usage of these substances is that ternary and quaternary compounds of these elements enable continuous tuning of the band gap by varying the composition, i.e the smooth tuning of the band-edge emission peak position.    
Synthesis of core-shell quantum dots with radial gradients of chemical composition is also possible on the basis of quaternary Cd$_{x}$Zn$_{1-x}$S$_{y}$Se$_{1-y}$ compounds, which enables variation of the exciton peak position in the range from 500 to 610~nm depending on the dot composition \cite{CdZnSSe_Single-step_synthesis_QDs,CdZnSSe_colloidal_QDs}.
Ternary Cd$_{x}$Zn$_{1-x}$Se nanocrystals also enable varying color of highly efficient exciton luminescence from blue to deep red demonstrating high stability \cite{Composition-Tunable_ZnxCd1-xSe,CdZnSSe_colloidal_QDs}.

Mention also, that Cd$_{x}$Zn$_{1-x}$S$_{y}$Se$_{1-y}$ thin films are known to be promising for utilization in quantum well lasers and solar cells \cite{CdZnSSe/ZnSSe_DLTS,CdZnSSe_green-Laser_diode,CdZnSSe_active_layers,CdZnSSe_active_layers_Book,CdZnSSe_PV-device}.

However, the band-edge luminescence of Cd$_{x}$Zn$_{1-x}$S$_{y}$Se$_{1-y}$ crystallites is weakly investigated and the luminescence of these crystals in glass as well as the luminescence of bulk CdZnSSe crystals, to our best knowledge, seems to be unstudied at present. 

Many of II--VI compounds are used as pigments both in art and technology \cite{Artists_Pigments_Cd,Pigment_Compendium}.
Nevertheless, we failed to find any mentioning of the CdZnSSe compound as a pigment in the literature including such exhaustive sources of data on pigments utilized in all areas of arts as ``Artists{\textquotesingle} Pigments'' \cite{Artists_Pigments_Cd} or ``Pigment Compendium'' \cite{Pigment_Compendium} despite that cadmium sulphides, selenides or even ternary compounds based on them, such as Cd$_x$Zn$_{1-x}$S, CdS$_x$Se$_{1-x}$, Cd$_x$Hg$_{1-x}$S,  were used in arts as pigments in the past and continue being employed at present.
The same should be said about pigments used in the industry.

\begin{figure}[t]
	\includegraphics[scale=0.7]{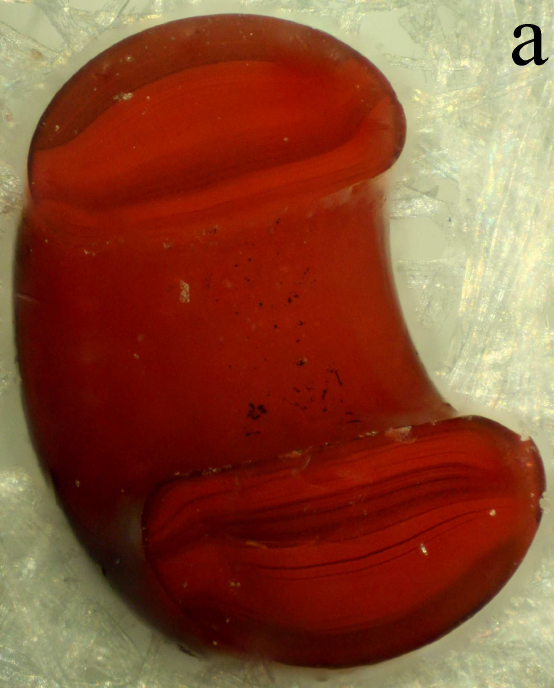}{\tiny{~}}%
	\includegraphics[scale=0.7]{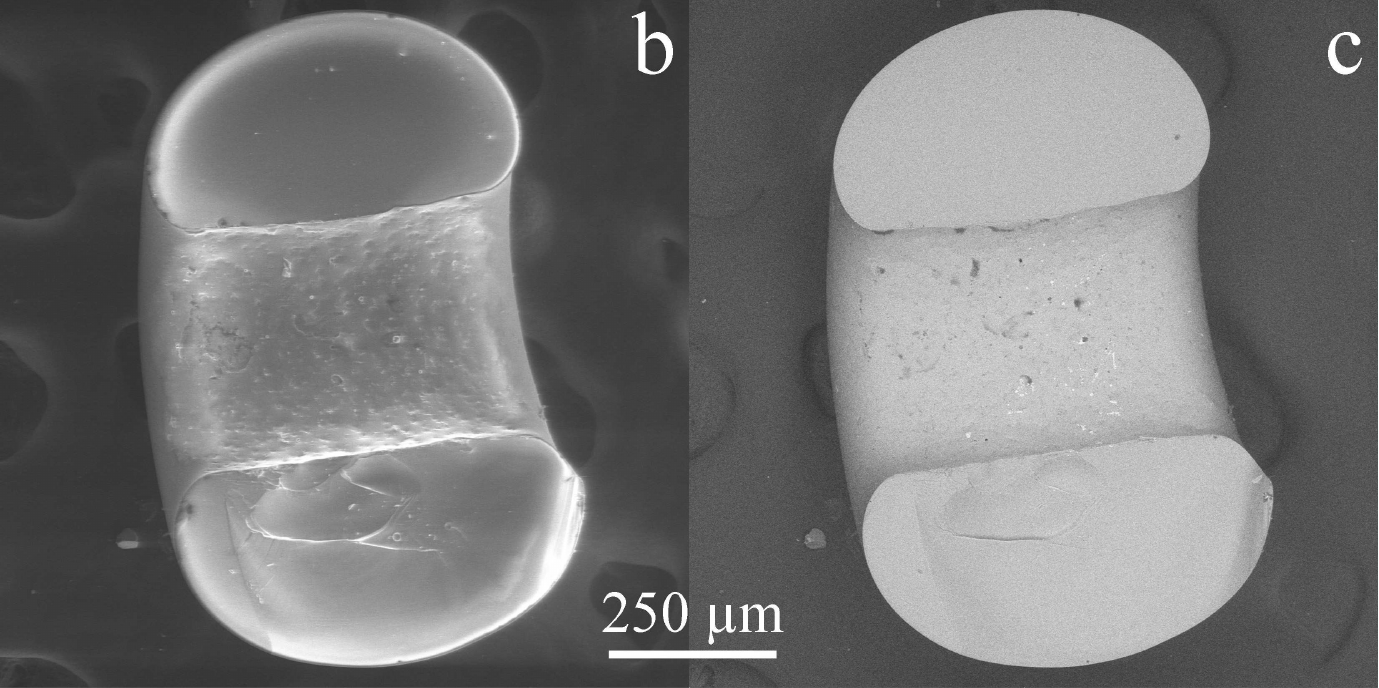}\\
	\includegraphics[scale=0.7]{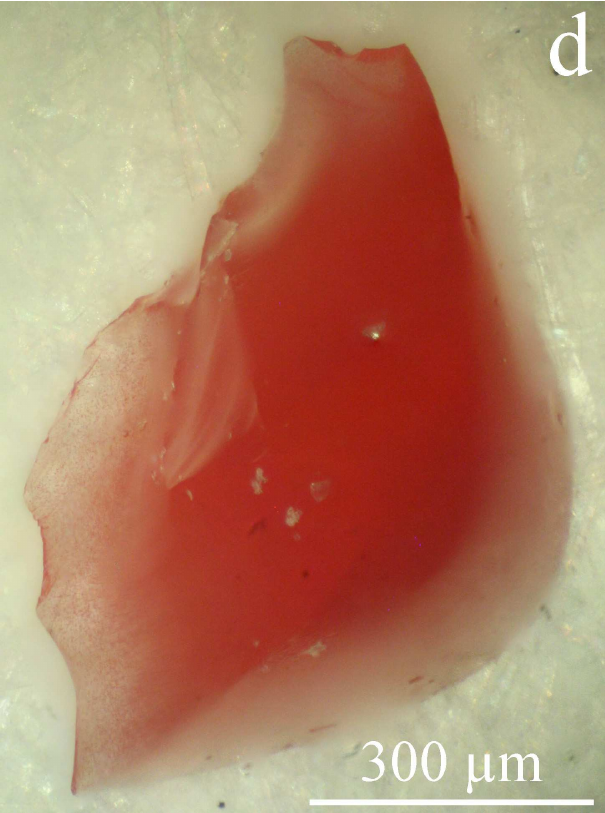}{\tiny{\,}}%
	\includegraphics[scale=0.7]{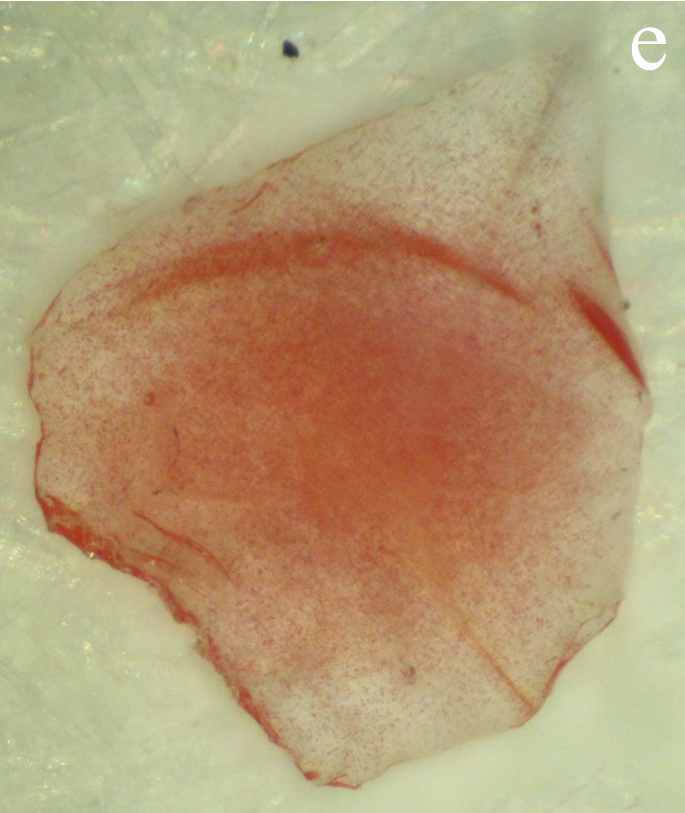}{\tiny{\,}}%
	\includegraphics[scale=0.7]{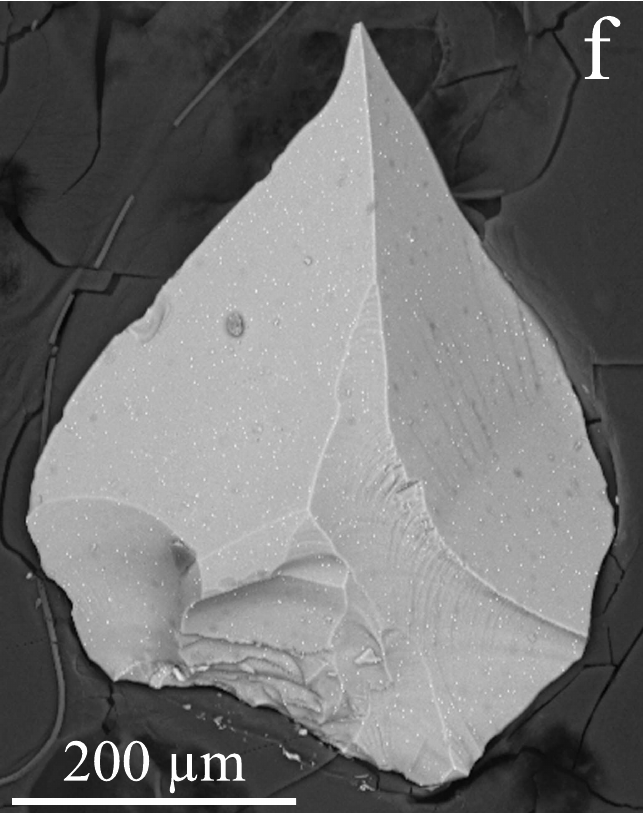}
	\caption{%
		Micro photographs and SEM images of some of the studied samples:  
		a micro photograph (a) of a half of a 19th century bead made of red 
		glass and SEM images of the same sample obtained in 
		secondary~(b)  
		and 
		backscattered~(c) electrons 
		at low vacuum;
		two pieces of red glass split off from a bead:
		the thicker one (d) and the thinner one~(e,\,f),
		panels (d,\,e) show micro photographs and (f) presents a SEM image in backscattered electrons; 
		red clusters producing colloidal staining of glass are seen in the 
		thinner sample~(e) and close to the sharp edge of the thicker one~(d); the clusters are also seen as tiny white spots in~(f).
		\label{fig:1}
	}
\end{figure}

Investigating glass corrosion in historical seed beads of the 19th century that often originates from crystalline inclusions synthesized in glass melt in the process of the glass production
\cite{KSS_Electron_microscopy,Yur_JAP,Yuryev_JOPT},
we have found out that Zn-rich red silicate glass made for the manufacture of seed beads contains CdZnSSe crystals \cite{Technart-2019_Crystals}.
Due to colloidal staining of glass, the CdZnSSe crystallites embedded in glass allowed bead manufacturers to obtain bright red glass (Fig.\,\ref{fig:1}) demonstrating, as we have ascertained, intense exciton luminescence from CdZnSSe inclusions that peaks around the wavelength of 610~nm \cite{CdZnSSe_CL_arXiv}.

It should be noted that CdZnSSe crystals\,---\,(CdS)$_x$(ZnSe)$_{1-x}$, $0<x<1$\,---\,are known to be synthesized and characterized using X-ray analysis only in the early 1960th \cite{CdZnSSe_structure}. 
Nevertheless, we have unexpectedly found the CdZnSSe crystals ($x=0.5$) in 19th century glass, which had been synthesized in it as a colorant. 

In this article, we investigate band edge cathodoluminescence (CL) spectra of CdZnSSe semiconductor crystallites (Cd$_{x}$Zn$_{1-x}$S$_{y}$Se$_{1-y}$, $x=y=0.5$) that nucleated and grew in glass melt during red glass beads manufacturing in the 19th century \cite{CdZnSSe_CL_arXiv}, determine the exciton emission energy and estimate the value of the band-gap width derivative ($dE_{\rm g}/dT$) in the interval from 80 to 300~K in CdZnSSe bulk crystals.

We also consider prospective applications of glass-crystal composite based on CdZnSSe and silicate glass in modern technologies and find it promising for utilization as bright red phosphor in glass (PiG) or as colored glass with CdZnSSe crystals serving as pigment particles for colloidal staining. 

The discovery of CdZnSSe semiconductor crystals in glass that was made in the 19th century is quite surprising.
This makes us discuss the issues concerning the possibilities of the commercial production of glass containing CdZnSSe crystallites in the 19th century such as the commercial availability of the components in the 19th century that might be used for making silicate glass stained with CdZnSSe crystals (Section~\ref{How?}).
These issues are undoubtedly mainly of the historical interest. 
However, we will try to clarify them in this article since the presence of such compound as CdZnSSe crystals used (especially intentionally used) as a pigment in glass of the 19th century seems quite unbelievable.
At the same time, the knowledge of processes and materials used for the synthesis of such material in the past may show the way for the creation of the modern technology of its production. 

Now, we proceed to a detailed presentation of the results of the study.

\begin{table}[t]
	\caption{%
		Mean relative content of some chemical elements in the examined glass samples 
		measured using XRF analysis (relative units proportional to at.\%).
		\label{tab:XRF}
	}
	\begin{tabular}{lccccccccccc}
		\noalign{\smallskip}\hline
		Chemical element & &Al      &Si     &Se     &Ca     &Fe     &Zn   	&As     &Pb     &S      &Cd\\
		\hline
		Mean content&      &1.50&	78.23&	0.39&	1.95&	0.04&	16.78&	
		0.45&	0.05&	0.14&	0.47\\
		Standard deviation&&0.34&	0.48&	0.01&	0.05&	0.01&	0.27&	
		0.02&	0.01&	0.07&	0.28\\
		\hline
	\end{tabular}
\end{table}

\section{Samples, Methods and  Equipment}
\label{exp}

\subsection{Sampling and Sample Preparation}
\label{sampling}

Glass specimens---bead fragments and small pieces of crumbled beads (Fig.\,\ref{fig:1})---were obtained during restoration of bead embroidered articles of the 19th century kept in museums.

Before all experiments, samples were washed with high purity isopropyl alcohol ([C$_3$H$_7$OH]~$>99.8$ wt.\,\%) at 40{\textcelsius} for 20 minutes in a chemical glass placed into an ultrasonic bath ($\nu=40$~kHz, $P=120$~W) \cite{Beads_KSS_SPIE}. 

All the samples were characterized using X-ray fluorescence analysis following the procedure described in Ref.\,\cite{KSS_Electron_microscopy}.

\subsection{Methods and Equipment}
\label{methods}

The following techniques and equipment were used in the research.

Elemental composition of glass was analyzed using a M4 TORNADO X-ray fluorescence (XRF) microspectrometer (Bruker).
The measurements were carried out following the procedure described in detail in Ref.~\cite{KSS_Electron_microscopy}.

Mira 3~XMU (Tescan Orsay Holding) scanning electron microscope (SEM) was employed for imaging of glass structure and elemental analysis. 
Mira 3~LMH (Tescan Orsay Holding) equipped with Nordlys Nano electron backscatter diffraction (EBSD) unit and AZtecHKL Advanced software (Oxford Instruments Nanoanalysis) was used for the structural analysis of crystallites in glass.

EDS X-MAX 50 (Oxford Instruments Nanoanalysis) energy dispersive X-ray spectrometers were used for elemental composition microanalysis and mapping.

Finally, micro cathodoluminescence (CL) was investigated both at room temperature and at liquid nitrogen temperature using MonoCL3 system (Gatan) mounted at JSM-7001F SEM (Jeol).

\begin{figure}[t]
	\includegraphics[width=0.4\textwidth]{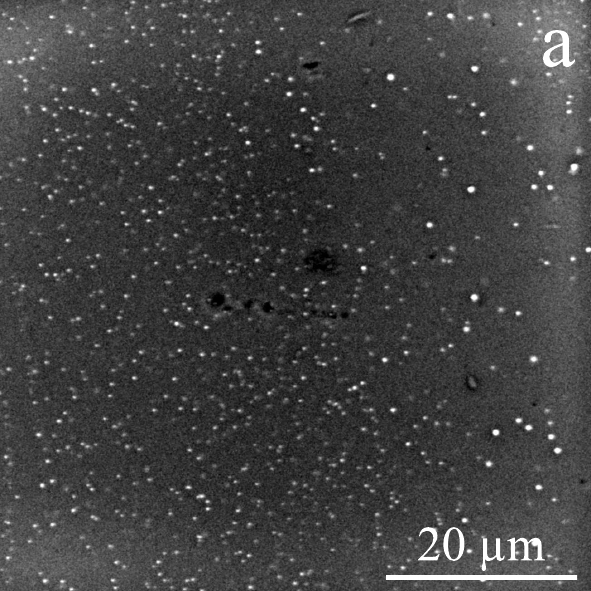}
	\includegraphics[width=0.4\textwidth]{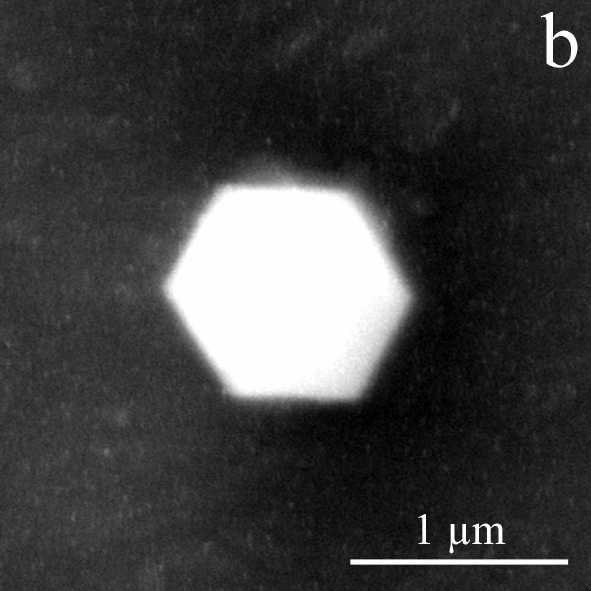}
	\caption{\label{fig:2}
		SEM images of CdZnSSe crystallites in glass: 
		(a) backscattered and (b) secondary electrons.
	}
\end{figure}

\begin{figure}[t]
	\includegraphics[width=\textwidth]{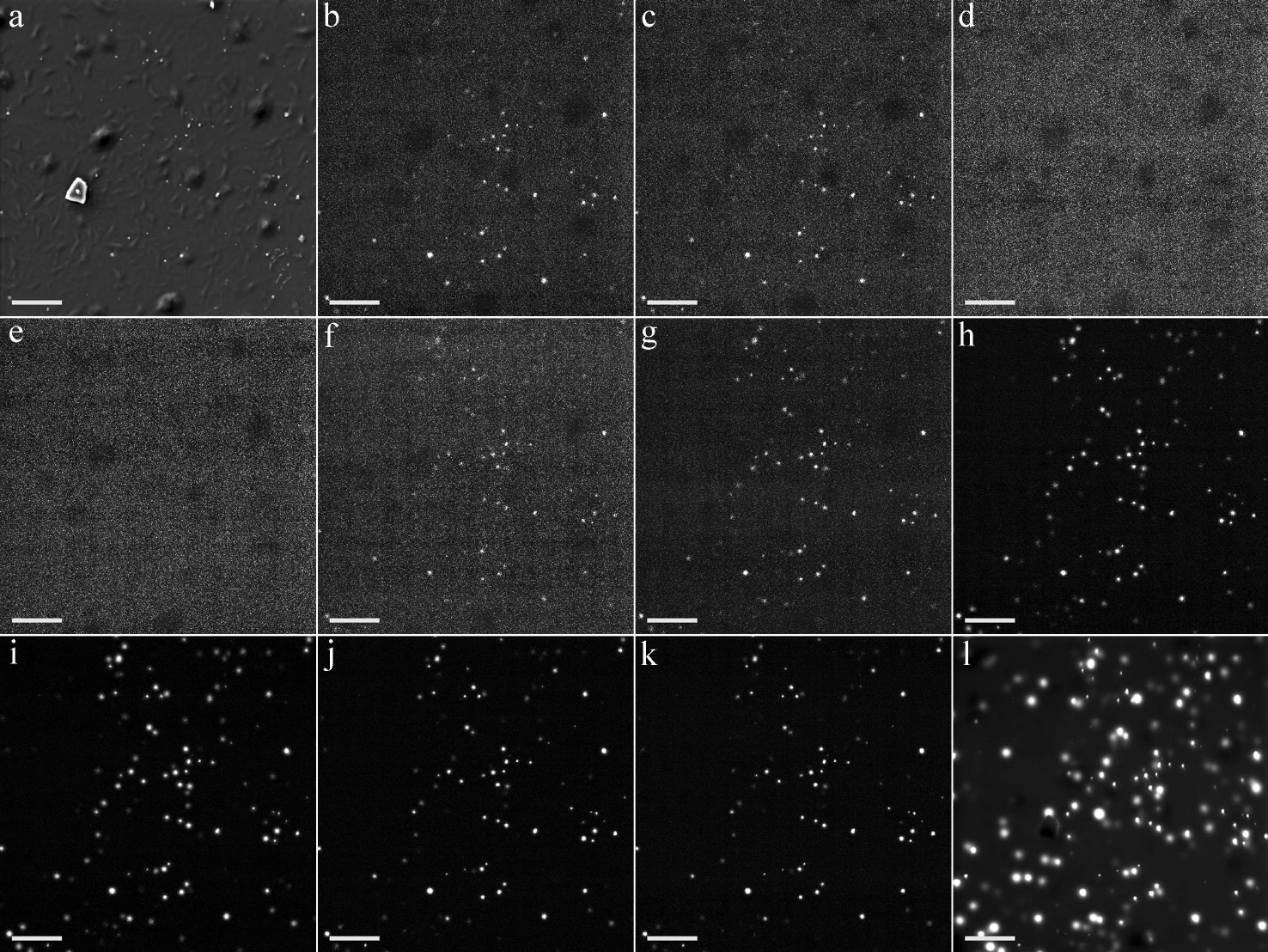}
	\caption{\label{fig:3}
		SEM SE image (a) of a red glass sample containing CdZnSSe crystallites, micro CL maps of the same area on the sample surface obtained at the wavelengths of  
		(b) 400, 
		(c) 410,
		(d) 430,
		(e) 505,
		(f) 560,
		(g) 570,
		(h) 590,
		(i) 610,
		(j) 635
		and
		(k) 640
		nm
		(the e-beam energy	$E_{\rm b}=10$~keV),
		and 
		(l) a panchromatic micro CL map of the same area obtained in the wavelength interval
		from 160
		to
		930~nm
		($E_{\rm b}=15$~keV); 
			the scale bar corresponds to 10 {\textmu}m on all the panels;
				$T = 300$~K.
			}
\end{figure}

\begin{figure}[t]
	\includegraphics[width=\textwidth]{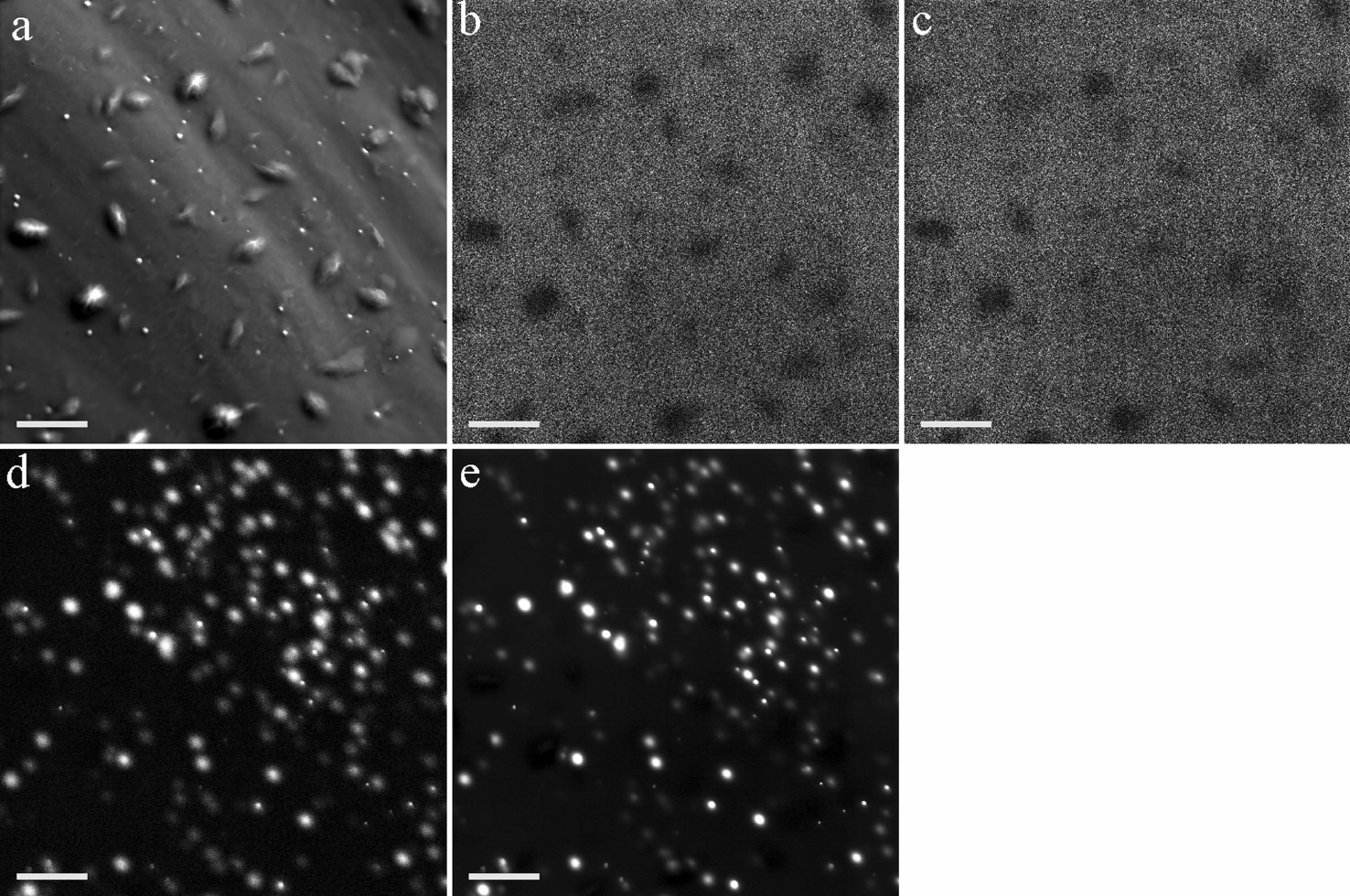}
	\caption{\label{fig:4}
		SEM SE image (a) of a red glass sample containing CdZnSSe crystallites, micro CL maps of the same area on the sample surface obtained at the wavelengths of  
		(b) 420, 
		(c) 475
				and
		(d) 602
				nm
		(the e-beam energy	$E_{\rm b}=15$~keV),
		and 
		(e) a panchromatic micro CL map of the same area obtained in the wavelength interval
		from 160
		to
		930~nm
		($E_{\rm b}=15$~keV); 
		the scale bar corresponds to 10 {\textmu}m on all the panels;
		$T = 80$~K.
	}
\end{figure}

\begin{figure}[t]
	\includegraphics[width=0.5\textwidth]{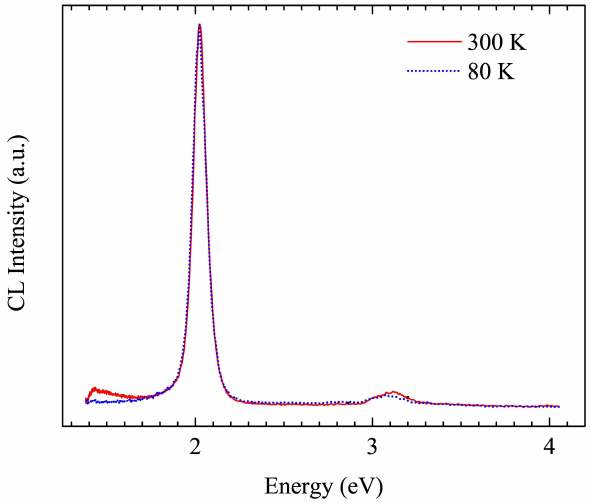}
	\caption{\label{fig:5}
		Survey spectra of micro CL recorded at 300 and 80~K at individual CdZnSSe crystallites 
		in glass.
	}
\end{figure}

\begin{figure}[t]
	\includegraphics[width=0.5\textwidth]{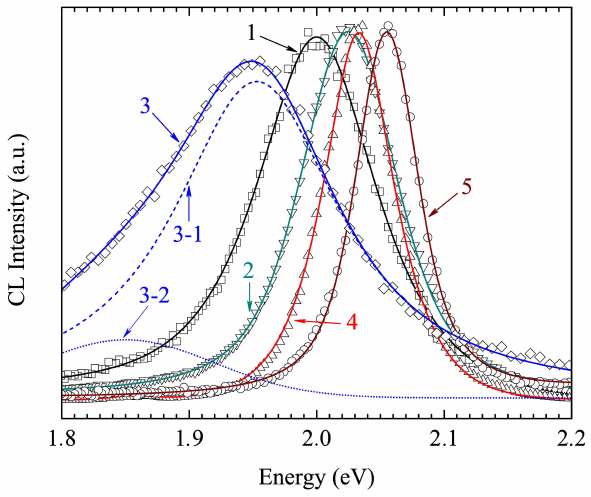}
	\caption{\label{fig:6}
		CL spectra of CdZnSSe crystallites in glass obtained at 300 (curves 1 
		to 3) and 80~K (curves 4 and 5); peaks 3-1 and 3-2 are components of  
		the band 3; Voigt profiles (lines) were used for fitting experimental 
		data (symbols).
	}
\end{figure}

\section{Results}
\label{res}

\subsection{Observation and identification of crystallites in glass}
\label{structure}

As a result of the composition and structure studies of turbid zinc-rich silicate glass of red seed beads using SEM (micro photographs and SEM images of some studied samples are presented in Fig.\,\ref{fig:1}, data of the glass XRF analysis are given in Table~\ref{tab:XRF}), we have found that it contains numerous sub-micron sized hexagonal crystallites (Fig.~\ref{fig:2}), which have been identified using EDS and EBSD as CdZnSSe (hexagonal crystal system, P6$_3$mc (186) space group; $a=4.052$\,{\AA}, $c=6.817$\,\AA)%
\footnote{Inorganic Crystal Structure Database (ICSD) \cite{ICSD,Crystallographic_Databases_article,Crystallographic_Databases} Entry~43368.} %
\cite{CdZnSSe_structure}.

This compound, to our knowledge, has never been observed in glass before.
It stains glass in bright red as a colloidal pigment.
Being a wide-bandgap semiconductor it demonstrates bright band-edge luminescence, like its binary counterparts, such as CdS, CdSe, ZnS and ZnSe, or ternary compounds on the basis of Cd, Zn, S, and Se in different combinations.

\subsection{CL maps and spectra}
\label{CL_spectra}

We have mapped panchromatic CL of a number of samples of analogous red glass from different museum articles,%
\footnote{That allows us to suggest that the studied samples of glass were made by different manufacturers or at different time in different batches, and consequently, their melting and vitrification conditions, as well as the conditions of thermal treatments at further technological processes during bead manufacturing \cite{KSS_Electron_microscopy}, were somewhat different.}
as well as CL at different wavelengths at the same places on the samples, and found the CdZnSSe crystallites to exhibit intense CL peaked around the wavelength of 610~nm, both at 300 and 80~K (Fig.~\ref{fig:3} and~\ref{fig:4} ).
The spatial distributions of particles radiating at the wavelengths around 610~nm and in the overall wide spectral range from 160 to 930~nm coincide.
Hence, we can state that mainly CdZnSSe crystals contribute to CL of this kind of glass and no other luminescing particles are present in this glass at appreciable quantities.

We also recorded CL spectra from individual CdZnSSe crystallites at 300 and 80~K.
Examples of micro CL spectra obtained directly from the crystallites are plotted in Fig.~\ref{fig:5}.
We assign the intense bands peaked around 2.0~eV to exciton in CdZnSSe crystals; the weak bands peaked in the range from 3.1 to 3.2~eV may be tentatively assigned to amorphous ZnO dissolved in Zn-rich glass \cite{BeZnO,a-ZnO,nano_ZnO_PL}.
The latter bands are related to CL of glass surrounding the crystals.
Objects emitting light in this spectral range are observed in the micro CL maps presented in Fig.~\ref{fig:3}\,a--c;
their spatial distribution correlates with that of CdZnSSe micro crystals.

Fig.~\ref{fig:6} presents the CL bands of  CdZnSSe in the vicinity of 2~eV obtained from different crystallites. 
Maxima of the CL bands are seen to shift for particular crystals: they usually move within the range from about 2.00 (curve~1) to nearly 2.03~eV (curve~2) at room temperature sometimes shifting down to 1.95~eV for some crystallites (curve 3-1).
The band peaked at 1.85~eV (curve 3-2) is likely due to the light admixture from glass surrounding the studied crystallites.  

No pronounced temperature dependence of peak maxima was observed for crystals of CdZnSSe. 
At liquid nitrogen temperature, the CL bands top out in the range from about 2.02 to nearly 2.06~eV (the spectral line shown by curve~4 peak at 2.03~eV and the line depicted by curve~5 reaches its maximum at 2.055~eV). 

We can roughly estimate the value of %
$dE_{\rm g}/dT$
in the interval from 80 to 300\,K  form the maximum temperature shift of the band edge CL peak position. 
The obtained estimate is %
$0>dE_{\rm g}/dT\gtrsim -2.5\times 10^{-4}$~eV/K.

\subsection{Stability of glass and crystallites}
\label{stability}

We have found that both glass and CdZnSSe crystallites are resistant to e-beam irradiation.%
\footnote{Sometimes, bead glass containing crystallites or crystallites themselves are unstable under the e-beam irradiation and decompose at the SEM beam power around $3.5\times 10^{-4}$~W and the absorbed power of about $1.5\times 10^{-5}$~W at room temperature yet some kinds of bead glass are highly resistant to e-beams of SEM or HRTEM \cite{KSS_Electron_microscopy}
}
We did not register any damages of glass or changes in the shape or the composition of crystals subjected to the SEM e-beam irradiation lasted for tens of minutes at room temperature and the absorbed power density of an order of 1~kW/cm$^2$.

The temperature and temporal stability of the crystals and glass as well as their resistance to long-term weathering will be discussed below (Section~\ref{Resistance}).
Mention only that we have not observed any signs of glass corrosion or degradation \cite{KSS_Electron_microscopy,Yur_JAP,Beads_KSS_SPIE} in the studied samples, which were prepared in the 19th century.

\section{Discussion}
\label{disc}

\subsection{Exciton emission, band gap variation, comparison with films and quantum dots}

As we have shown above (Section~\ref{CL_spectra}), the exciton CL band of CdZnSSe crystallites in glass peaked in the vicinity of 2~eV demonstrates slightly different maxima for separate crystals. 
The exciton band maximum varies in the interval from about 2.02 to nearly 2.06~eV at 80~K and from about 2.00 to nearly 2.03~eV at 300~K.
In our opinion, this  observation may be understood if minor variations in the elemental composition of the crystallites are considered, which may occur due to local fluctuations of the glass melt elemental composition during the synthesis.

The bandgap of CdZnSSe in the form of films or quantum dots has already been studied in the literature.

The value of optical band gap of polycrystalline CdZnSSe films was previously presented in Ref.~\cite{CdZnSSe_PV-device}. 
In that research, cadmium zinc sulphoselenide films were prepared using chemical bath deposition on glass substrates. 
The film band gap estimated from light absorption spectra was shown to decrease due to annealing at the temperatures of 100, 300 and 500{\textcelsius} from 2.21 to 2.01~eV with the increase of annealing temperature in that work. 
According to measurements of the X-ray diffraction the polycrystalline films demonstrated pure hexagonal structure after annealing.
The authors of that article explained such a behavior of the band gap by the decrease in originally increased interatomic spacing in the crystal grains or by particle size enlargement and enhancement in crystallinity with increasing annealing temperature.
This assumption looks plausible in the case of polycrystalline films on glass. 
However, we believe that shifts in the exciton CL band maxima, which we observe in CdZnSSe micro crystals encapsulated in glass that are studied in this work, cannot be explained by those factors.
First of all we should take into account the level of knowledge and technology development in the mid 19th century when the studied glass samples were made.
Evidently, no one of glass makers made any efforts to achieve high uniformity of  crystalline properties, their perfection or strict adherence to stoichiometry at the time of production of glass that we study.
Moreover, no one knew their chemical composition or had an idea about the structure of the crystals. 
The inclusions grew in glass just like ``wild'' natural crystals in minerals.
Nevertheless, all the crystallites were synthesized in glass melt at similar conditions at high temperature,
much higher than the annealing temperatures  in Ref.~\cite{CdZnSSe_PV-device}, 
certainly
higher than 1000{\textcelsius}
and likely 
close to 1500{\textcelsius},%
\footnote{We do not know exactly the temperature, at which this kind of glass was melted and bubbled in the 19th century, yet we can estimate its melting temperature from the melting point of the ZnO{\textperiodcentered}SiO$_2$ composition that is 1437{\textcelsius} according to Ref.\,\cite{Kitaygorodsky}.}
and although their dimensions vary in the range from several hundreds of nanometers to about one micrometer their lattice parameters measured at numerous individual crystals using EBSD are equal, thus neither the variation of their sizes nor the difference in the crystal perfection may explain the observed shifts of the exciton peak.
We explain this observation by minor changes in elemental composition of crystallites, which cannot be registered by EBSD within the accuracy of experiments yet can slightly change the band gap.
For example, the stoichiometry shift towards CdSe would decrease the band gap, whereas composition changes towards CdS, ZnSe or ZnS would increase it.
Hence, minor deviations of the CdZnSSe elemental composition from strict 
Cd$_{0.5}$Zn$_{0.5}$S$_{0.5}$Se$_{0.5}$
would result in red or blue shifts of the exciton luminescence peak.
It should be noted also that the same hexagonal type of lattice and equal lattice parameters ($a=4.052$\,{\AA}, $c=6.817$\,\AA) were given for crystals with the 50\%\,CdS--50\%\,ZnSe composition and the 40\%\,CdS--60\%\,ZnSe one in the earliest known by us publication on the structure of CdZnSSe compounds (Ref.\,\cite{CdZnSSe_structure}), on which the corresponding ICSD Entry is based.
This means that one cannot distinguish the quaternary compounds belonging to the above interval of crystal compositions using electron or X-ray diffraction techniques.   
Thus, if the composition of the studied crystals varies in different precipitates within this interval we could not register such variations by EBSD but could observe shifts in exciton CL peak position. 

Another article cited in Ref.\,\cite{ZnSe_0.5CdS_0.5} also presents data on the band-edge luminescence of CdZnSSe. 
(ZnSe)$_{0.5}$(CdS)$_{0.5}$ films with the thickness of about 600~nm were deposited on glass substrates at the temperatures of 350 and 470~K by thermal vacuum evaporation from the sources of a mixture of ZnSe and CdS sintered at 1100{\textcelsius} for about 48 hours in vacuum-sealed quartz tubes in that work.
The resultant polycrystalline (ZnSe)$_{0.5}$(CdS)$_{0.5}$ films obtained at different temperatures were found to have the wurtzite structure but different lattice parameters: $a=4.048, c=6.565$\,\r{A} and $a = 4.062, c = 6.605$\,\r{A} for films grown at 350~K and 470~K, respectively, which are obviously different from those given in the original article that is used in ICSD \cite{CdZnSSe_structure}.
Optical band gaps of those films derived from their transmission spectra were also slightly different (2.43 and 2.45~eV at room temperature for films formed at 350~K and 470~ K, respectively).
Grain sizes were 371 and 574\,\r{A} for the films obtained at deposition temperatures of 350~K and 470~K, respectively, that obviously results from the increased mobility of the film components at the higher temperature. 
The difference in structural parameters and the band-gap width were explained by the authors by the deviations of the film composition from 25\% for each element, which were considerably lower in the films grown at 470~K.
According to Ref.\,\cite{ZnSe_x-CdS_1-x_films}, the lattice parameters of (ZnSe)$_{x}$(CdS)$_{1-x}$ films grown at 470~K vary linearly with $x$ following Vegard's law that means that films of these alloys are miscible in the entire range of the composition ($0\leqslant x \leqslant 1$).
Yet, the optical band gap varies non-linearly demonstrating concave dependence on $x$ with a minimum at $x=0.2$ ($E_{\rm g} \approx 2.4$~eV).

Unfortunately, the results of Refs.\,\cite{ZnSe_0.5CdS_0.5} and \cite{ZnSe_x-CdS_1-x_films} related to the band gap variations with $x$ cannot explain the drastic difference in the band gap presented in those articles and the exciton peak position registered in this work. 
To explain it we can assume that this discrepancy is caused by the difference in the lattice parameters of the crystals synthesized in glass melt at high temperature and polycrystalline films deposited at much lower temperatures on glass substrates. 
The liquid environment, in which the initial components were dissolved, implies the higher miscibility of the elements during the synthesis and the enhanced homogeneity of the distribution of atoms in the resultant CdZnSSe micro precipitates.
Strains that were likely present in grains constituted the CdZnSSe films on glass in Refs.\,\cite{ZnSe_0.5CdS_0.5} and \cite{ZnSe_x-CdS_1-x_films} should also be taken into consideration.
In the case of chemical synthesis of CdZnSSe micro crystals in glass melt, no strains should be expected or, if occur, they should be much less  than in films deposited on glass, especially taking into account the absence of fractures in glass, which arose neither at the crystals nor far from them during the melt vitrification, glass cooling, its repeated thermal treatments in the process of bead making or during the further usage of glass beads for more than a century.%
\footnote{%
	Fractures usually arise in glass if it is highly stressed because of the presence of crystallites with the temperature linear expansion coefficients strongly different from that of glass like it takes place in turquoise glass containing crystalline inclusions of orthorhombic KSbOSiO$_4$ \cite{KSS_Electron_microscopy}. 
	In that case crystallites should also be under the mechanical stress.
}
The latter suggests that glass and likely the crystalline inclusions are virtually unstrained.

Photoluminescence (PL) spectra resembling the spectra shown in Fig.\,\ref{fig:6} were obtained from Cd$_{x}$Zn$_{1-x}$S$_{y}$Se$_{1-y}$ core/shell quantum dots with chemical composition gradients \cite{CdZnSSe_Single-step_synthesis_QDs,CdZnSSe_colloidal_QDs}.
Those chemically synthesized quantum dots had a zinc blende structure; their composition monotonically varied along the radii.  
By varying the initial precursor composition of the dots synthesis reaction the authors of Ref.\,\cite{CdZnSSe_Single-step_synthesis_QDs} succeeded to control the exciton PL peak position of quantum dots shifting its maximum wavelength between 500 and 610~nm.
The most red-shifted peak reported in that work corresponds to the CL peaks shown in Figs.\,\ref{fig:3} and~\ref{fig:6}.
The similar properties of Cd$_{1-x}$Zn$_{x}$Se core-shell quantum dots were reported in Ref.\,\cite{Composition-Tunable_ZnxCd1-xSe}.
With increasing Zn content, a composition-tunable emission across nearly the entire visible spectral range had been achieved by a controlled blue shift in the emission wavelength.
The authors of those works especially emphasized the stability of quantum dots that they synthesized.
Concerning crystals in glass discussed in this article, it is also possible to synthesize Cd$_{x}$Zn$_{1-x}$S$_{y}$Se$_{1-y}$ crystallites of the controlled composition and obtain CdZnSSe-based glass-crystallites composites with tunable emission in a wide spectral range by varying the atomic fractions of Zn, Cd, Se and S in the glass melt.
The stability of the crystallites in glass is also very high.
Being encapsulated in glass the studied inclusions were kept for more than a century without any visible changes in their color.  
They are highly luminescent.
Additionally, Zn-rich glass, in which they are embedded, also demonstrates a very high stability; it is also highly resistant to the e-beam irradiation and atmospheric moisture. 

\subsection{CdZnSSe as phosphor in glass}
\label{PiG}

It is evident from the above that CdZnSSe in zinc glass may serve as an efficient and stable light converter suitable for use, say, in light emitting diodes (LEDs) as a replacement for presently used silicone phosphors, i.e. they are promising as phosphor in glass (PiG) \cite{PL_Glasses} especially taking into account the challenge in luminescent performances of PiG that exists because of the lack of PiG emitting in red \cite{PiG_red-emitting,PiG_white_LED}.

Recently, the authors of Ref.\,\cite{PiG_red-emitting} introduced 3.5MgO{\textperiodcentered}0.5MgF$_2${\textperiodcentered}GeO$_2$:Mn$^{4+}$ (MFG:Mn$^{4+}$) phosphor particles into a 55TeO$_2$--25ZnO--16Na$_2$O--4Al$_2$O$_3$ (in mol\,\%) glass matrix to obtain chemically stable PiG emitting in deep red. 
Combining MFG:Mn$^{4+}$ with the commercial YAG:Ce$^{3+}$ phosphor in tellurite glass matrices enabled the demonstration of a white laser diode with tunable colour temperature (CCT) \cite{PiG_white_LD}.
Alternatively, La$_2$Ti$_2$O$_7$:Eu$^{3+}$ bright red phosphor was synthesized and introduced together with commercial YAG:Ce$^{3+}$ one in TeO$_2$--ZnO--Sb$_2$O$_3$--Al$_2$O$_3$--B$_2$O$_3$--Na$_2$O glass that enabled obtaining in combination with InGaN blue chip a high-power white LED with CCT of 4809~K.
The commercial CaAlSiN$_3$:Eu$^{2+}$ (CASN:Eu$^{2+}$) red nitride phosphor embedded into an ultra-low melting Sn--P--F--O glass together with YAG:Ce$^{3+}$ allowed one to obtain bright warm white LED with CCT of 3554~K \cite{PiG_tunable}.
Patterned PiG with separated red CASN:Eu$^{2+}$ and yellow YAG:Ce$^{3+}$ segments was obtained by screen-printing followed by low-temperature sintering \cite{PiG_screen-printing}.

The given examples demonstrate the usage of rare-earth based phosphors for obtaining efficient PiGs emitting in red. 
All of them require external synthesis of the phosphors followed by sintering in glass charge and low-temperature glass melting.
CdZnSSe crystallites in zinc glass are synthesized directly in the melt that is an evident advantage for the production on the commercial scale.
In addition, the process of cadmium glass making is well developed and available at numerous glassworks throughout the world.
All the components required for this process are commercially available including such compounds as CdS, CdSe, ZnO, ZnS or ZnSe.
By varying the fractions of these components one can vary not only the glass tint but its color from blue to red that is also attractive for producing light sources of different colors or tints of white.

Thus, we consider the CdZnSSe--glass composite as a promising highly stable and temperature resistant PiG with the tunable color.%
\footnote{It is also resistant to the e-beam irradiation at least up to the energies of 30~keV.
}

\subsection{Resistance to e-beam irradiation, heating and weathering; temporal stability}
\label{Resistance}

We write in Section~\ref{stability} that both glass and CdZnSSe crystals are resistant to the SEM e-beam irradiation.
This is an important feature for luminescent glass.


In this section, we will discuss other properties, which are also important for the practical applications of this kind of glass, such as its stability during heating, the resistance to weathering and temporal stability.
We did not specially investigate these issues experimentally, yet the required conclusions can be made from the bead manufacturing technology. 
Long history of being of this kind of glass in beaded articles also gives an information about its temporal stability and resistance to weathering and some other external factors such as, e.g., effect of detergents or sebum.

It is known that the process of bead manufacturing comprised a number of steps such as the production of a glass tube including its drawing until the diameter reached 1 mm, tube cutting to small pieces, usually less than 1 mm in length, followed by tumble finishing that enabled obtaining brilliant glass seed beads with smooth edges
\cite{GLASS_AN_USSR,KSS_Electron_microscopy,Yurova_large}.
Each of these steps was conducted at an elevated temperature specific for the certain process and was terminated by glass cooling at some rate often not precisely controlled.
Tumbling finishing at elevated temperatures often introduced structural inhomogeneity and internal stress in glass of seed beads that in turn launched long-term fracturing of glass \cite{KSS_Electron_microscopy,Beads_Micro-FTIR_Kadikova_article,RCEM-2018_vibrational_spectroscopy}.
Crystallites sometimes played a key role in arising of internal stress fields \cite{KSS_Electron_microscopy}.

Glass beads studied in this article were also produced using the above process. 
However, they demonstrate an excellent preservation state in museum exhibits due to zinc glass composition and structure.
This fact demonstrates that studied glass is highly resistant to annealing including thermal cycling (unless it is reheated to near the softening point at which its colour possibly might become deeper \cite{Se_volatility_Glass_1929}). 
Additionally, it demonstrates the long-term temporal stability and proves that this kind of glass is not subject to slow degradation caused by the internal stress.
High melting temperature of this kind glass and small sizes of crystallites, which do not form colonies generating internal stress in glass, facilitate the resistance to thermal treatments and the long-term stability.

Red glass studied here is highly resistant to the effect of weathering or atmospheric moisture.%
\footnote{%
	Note also that glass studied in this article is stable to effect of soap, detergents and other cleansing agents, which were used for washing beaded articles and cleaning beads initially during the domestic use and later during museum keeping.
	It is also resistant to sebum that undoubtedly contacted with beads when the beaded items were in home use.
}
It has not degraded and has remained undamaged for more than a hundred years of it life.
That is why we consider this glass as a perfect medium for the encapsulation of embedded CdZnSSe crystals.
Glass prevents crystal surface oxidation by atmospheric oxygen that otherwise might be considerable for a long time period, say, as in the considered case, during the time exceeding a hundred years \cite{II-VI_surface_oxidation,Poy-ZnSe_oxidation_@430-700C,Poy-ZnSe_surface_oxidation}.

Crystals encapsulated in glass are also well preserved against effects of annealing.
Glass surrounding crystal keeps them from losing volatile components, which could leave the crystals during thermal treatments at other conditions \cite{Cd_volatility,ZnSe:82Se}.

Thus, we can conclude now that composite materials based on microcrystallites and nanocrystallites of II--VI compounds embedded into silicate glass may find many technical applications in photonics due to their bright luminescence and high stability and resistance to external influences.

\subsection{CdZnSSe as glass pigment}

As for pigments, as mentioned above, we failed to find any information concerning such applications of CdZnSSe.
Ternary II--VI compounds---various Cd sulpho{\-}sele{\-}ni{\-}des---providing orange and red shades have been used as art pigments since 1910 \cite{Artists_Pigments_Cd,Pigment_Handbook,Pigment_Handbook_2nd}.
Presently, they are commercially available.
Apart from that, CdSe$_{0.33}$S$_{0.66}$ suspended in glass is used for staining red glass utilized for railway, marine and other signaling lights \cite{Artists_Pigments_Cd}.
The data presented in this article, however, gives an evidence that CdZnSSe was employed already in the 19th century for staining glass used for manufacture of seed beads.
It turned out that the process of its chemical synthesis in glass melt was developed already at that time that enabled producing brilliant glass of a lovely red tint.
It is evident, however, that the process was found empirically since the chemical composition and structure of this colloidal pigment was undoubtedly unknown to glass makers.

\subsection{On the possibility of CdZnSSe crystallites synthesis in bead glass in the 19th century and the glass production time}
\label{How?}

The issue of possibility of CdZnSSe crystallites synthesis in commercial bead glass in the 19th century is intriguing from both technological and historical viewpoints. 
Here, we will discuss whether it was possible, how it could be done and, finally, when glass containing CdZnSSe crystals could be made.

As we have already mentioned above, CdZnSSe crystals formed in glass melt at high temperature.
However, it is unclear what initial components were added into the charge to obtain the CdZnSSe crystals.

ZnO was certainly used to make zinc glass.
Although this compound has been known since antiquity, its production on an industrial scale and therefore the suppositional usage as a glass pigment might be started in 1845 in France and in the 1850s in other European countries (and the USA) \cite{Artists_Pigments_Zn-White}.
Thus, the studied glass is unlikely to have been produced before those years.

An issue of the sources of Cd, Se and S is much more complicated.

Sulphur might be added as an individual substance.
However, it seems more likely that it was added together with Cd in the form of CdS that was synthesized by Gay-Lussac in 1818 and suggested as an artists{\textquotesingle} pigment by Stromeyer in 1819 \cite{Artists_Pigments_Cd}.
Metallic Cd was being produced in Upper Silesia in 1829; but the commercial CdS pigment (cadmium yellow) has been produced since the mid-1840s \cite{Artists_Pigments_Cd}.

Despite the fact that selenium was discovered by Berzelius as an admixture of sulphur in 1817, the usage of Se-based pigments, such as CdSe or Cd sulfoselenide, had not been mentioned in the literature until 1892 when cadmium sulfoselenide was patented \cite{Artists_Pigments_Cd,Painting_Materials,Se_ruby-1892}.
According to Refs.~\cite{Artists_Pigments_Cd,Painting_Materials},
Cd sulfoselenide was likely commercialized only about 1910 and the production of selenium red glass on its basis was reported in 1919 \cite{Selenium_Glass}.%
\footnote{%
	Although Krak states in Ref.~\cite{Se_volatility_Glass_1929} citing the patents of 1892 and 1894 \cite{Se_ruby-1892,Se_ruby-1894} that the first record on selenium ruby was made in 1891.	
}
Kirkpatrick and Roberts synthesized two types of red glass---a soft-working zinc-alkali glass and a plate glass \cite{Selenium_Glass}.
Batches for the first one contained ZnO, CdS and Se; those for the second one did not contain ZnO.
The glass melting temperature was 1400{\textcelsius} and was never risen in a lehr or a glory hole in excess of this temperature (see Ref.\,\cite{Selenium_Glass} for the detailed description of the processes used and the glass color at each production step).%
\footnote{%
	The given melting temperature corresponds well with our estimate of the studied bead glass melting temperature presented above. 
}
It is important to note that the glass color considerably changed at different production steps due to thermal treatments at different temperatures and durations being only lightly tinged with color at the early step and becoming deeper with successive heatings and coolings.
It became deep red as a result of the complete set of treatments.
The structure of these glasses is unknown to us.
Yet, we can suppose taking into account the notes concerning the possible formation of some crystallites in glass found in the cited article that CdSSe crystals formed in the plate glass and CdZnSSe ones arose in the zinc-alkali glass although the resultant glass, according to the diagrams plotted in that article, was deeper red than glass studied by us that may be due to admixtures of different II--VI and other compounds dissolved in those glasses. 
The authors of that article remarked in the conclusion that the composition of such glasses had been known for a considerable length of time.
Thus, following that remark we can assume that the recipes of similar selenium glasses might be known to bead makers of the 19th century.
The only issue is if they could obtain selenium in commercial quantities.
According to Ref.\,\cite{Legend_Bohemian_Glass}, selenium has been used in glass making since the end of the 19th century in Bohemia.
So, it might have been utilized in making bead glass for the production of Bohemian ``perles''.
More accurate dating of this glass is difficult.
We know, however, that Se was being investigated in labs in the 1870s (e.g., by Smith in 1873 or by Adams and Day in 1876) when its photosensitivity was discovered by Smith \cite{Action_Light_Selenium}.
Hence, this substance might be available in commercial amounts already in the 1870s or the 1880s.
Thus, it might have been employed for bead glass production since that time, and this allows us to attribute the studied glass beads to that time.
This dating is confirmed by the attribution of the museum exhibits, from which the fragments of seed beads were obtained during the restoration---they are attributed to the second half of the 19th century when beads of bright red color have appeared in beadworks \cite{Yurova,Yurova_large}.
The dimensions of the beads also may be considered as evidences in favor of this dating: the bead sizes are typical for the mid-19th century and less typical for the late 19th century but these seed beads are too small for the early 20th century.

Thus, we date the beginning of the manufacture of glass containing crystalline inclusions of CdZnSSe red pigment to the last quarter of the 19th century.

\subsection{Is Cd$_{0.5}$Zn$_{0.5}$S$_{0.5}$Se$_{0.5}$ thermodynamically equilibrium?}
\label{termodinamic}

It is interesting to know whether hexagonal Cd$_{0.5}$Zn$_{0.5}$S$_{0.5}$Se$_{0.5}$ is a thermodynamically equilibrium composition or it is a kinetic product of the process of its synthesis in the glass melt.

Since Zn was contained by a large excess compared to Cd, S and Se in the glass melt during the synthesis of CdZnSSe crystals (Table~\ref{tab:XRF}) and the latter elements were likely contained in different concentrations in the charge, probably not very accurately controlled, it can be concluded that the obtained crystals have thermodynamically equilibrium structure and composition.
Otherwise individual crystals would have strongly different compositions because of the inhomogeneous distribution of Cd, Se and S in the glass melt that follows from the non-uniform distribution of CdZnSSe crystals observed in Figs.\,\ref{fig:2}\,d,\,e and~\ref{fig:2}\,a.
Additionally, the CdZnSSe crystals would demonstrate variations in the stoichiometry in glass samples obtained from different beads and different museum articles since this glass was highly likely produced by different manufacturers that used somewhat differing production processes and conditions.

Thus, we should conclude that the hexagonal 
Cd$_{0.5}$Zn$_{0.5}$S$_{0.5}$Se$_{0.5}$ 
crystalline phase in red zinc glass is the thermodynamic product of the synthesis process in question.

\section{Conclusion}
\label{concl}

In summary, we would like to emphasize the main statements of the article.

We have observed hexagonal crystallites of sub-micron sizes in 19th century Zn-rich seed bead glass and identified them as CdZnSSe crystals ([Cd]\,= [Zn]\,= [S]\,= [Se]\,= 0.25).

The crystallites exhibit intense exciton cathodoluminescence both at room temperature and at liquid nitrogen temperature.

Exciton CL of these crystals tops out in the range from 2.00 to 2.03~eV at room temperature and from 2.02 to 2.06~eV at liquid nitrogen temperature.
Shifts of the band edge CL peak position in different crystallites are supposedly due to minor variations of stoichiometry of individual crystallites that lead to changes in the band gap width.

Finally, the temperature dependence of the band-edge CL peak position is weak or even negligible in CdZnSSe crystals.
We have estimated the value of $(dE_{\rm g}/dT)$ in the interval from 80 to 300~K; the obtained value ranges from 0 to $-2.5\times 10^{-4}$~eV/K.

We conclude also that the hexagonal Cd$_{0.5}$Zn$_{0.5}$S$_{0.5}$Se$_{0.5}$ crystalline phase in red zinc-rich silicate glass is a thermodynamic product of the synthesis reaction.

It should be noted in conclusion that luminescent glass-crystal composites or ceramics based on Zn or Cd-rich silicate glasses, similar to that made in the late 19th century and presented in this article, due to their optical properties and high stability and resistance to external influences may find numerous practical applications in photonics. 
Varying glass melt composition and crystal growth conditions to obtain crystallites of the required II--VI compound with the required sizes and in the required concentration, one can produce luminescent glasses with the required optical properties, which are easy to shape either using standard technologies of making optical parts or glassware  or using the 3D printing technology \cite{3D_printing_ceramics}.
The latter makes such materials attractive to be used also in arts and crafts.

Cd$_{x}$Zn$_{1-x}$S$_{y}$Se$_{1-y}$--glass composites are also promising as highly stable and temperature and e-beam resistant phosphors in glass with the color tunable from blue to red applicable for light conversion or as scintillators.

\section*{Acknowledgments}

We thank Miss Darya Klyuchnikova of GosNIIR for the preparation of the samples to the experiments. 
We also grateful to Ms. Lyubov Pelgunova of A.~N.~Severtsov Institute of Ecology and Evolution, the Russian Academy of Sciences, for the XRF analyses.
We express our appreciation to Ms. Ekaterina Morozova of GosNIIR 
for fruitful discussions of the results.

This work was supported by the Russian Science Foundation through the grant number 16-18-10366 and the Russian Foundation for Basic Research through the grant number 18-312-00145.
The research was accomplished under the collaboration agreement between {GosNIIR} and GPI RAS.

Mr. Sergey Malykhin is a grantee of the ``BASIS'' Foundation; he also appreciates the support from the Academy of Finland via the grant number 298298.%
\\

{\flushleft\textbf{Declarations of interest:} none.}


\section*{References}

\bibliography{Beads}

\end{document}